\documentclass[pdftex,twocolumn,epjc3]{svjour3}          

\RequirePackage[T1]{fontenc}

\smartqed  

\RequirePackage{graphicx}
\RequirePackage{mathptmx}      
\RequirePackage{flushend}
\RequirePackage[numbers,sort&compress]{natbib}
\RequirePackage[colorlinks,citecolor=blue,urlcolor=blue,linkcolor=blue]{hyperref}

\journalname{Eur. Phys. J. C}

\begin{document}
	
	\title{A unified thermodynamic framework for
		coextensive dark matter admixed strange stars}
	
	
	\author{Samstuti Chanda\thanksref{e1,addr1,addr2}
		\and
		Ranjan Sharma\thanksref{e2,addr1} 
	}
	
	\thankstext{e1}{e-mail: schanda93.dta@gmail.com}
	\thankstext{e2}{e-mail: rsharma@associates.iucaa.in (corresponding author)}
	
	\institute{IUCAA Centre for Astronomy Research and Development (ICARD),
		Department of Physics, Cooch Behar Panchanan Barma University,
		Vivekananda Street, Cooch Behar, 736101, West Bengal, India.\label{addr1}
		\and
		\emph{Present Address:} Department of Physics, Islampur College, Islampur, 733202, West Bengal, India\label{addr2}
	}
	
	\date{Received: date / Accepted: date}

	\maketitle
	
	\begin{abstract}
		We investigate the structural and physical properties of a strange star admixed with self-interacting bosonic dark matter. The total energy density is modelled as a weighted combination of quark matter and dark matter components regulated by a fixed local volume fraction. The quark component is described by a linear equation of state, while the dark matter follows a mean-field EOS with repulsive self-interactions. By combining these EOSs into a barotropic effective EOS derived from a unified thermodynamic potential, the two-fluid system is reformulated as a thermodynamically closed and mechanically equilibrated configuration. The construction preserves the dynamical distinction between the quark and dark sectors but treats them as a macroscopically unified mixture governed by a single hydrostatic equilibrium equation. This framework identifies the entirely coextensive limit of two-fluid models as a physically meaningful and thermodynamically closed configuration, providing a coherent macroscopic closure that links dark matter-strange matter microphysics to stellar observables. Using the effective EOS, we solve the governing Tolman-Oppenheimer-Volkoff (TOV) equations to obtain the mass-radius relationship by varying the model parameters. Our results reveal distinct modifications to the $M-R$ profiles, suggesting observable signatures that could offer insights into the impacts of dark matter in extreme astrophysical environments. We note that even a modest dark matter admixture softens the effective equation of state and shrinks the maximum mass limit. We discuss the relevance of our investigation in the context of recent observational data available for pulsars, such as XTE~J1814-338, PSR~J0348+0432, PSR~ J0740+6620, and PSR~J0952-0607.
	\end{abstract}
	
	\section{Introduction}
	
	Cosmological observations indicate that the universe is composed of approximately 68\% dark energy, 27\% dark matter (DM), and only 5\% ordinary baryonic matter (BM). The observational evidences stemming from the Cosmic Microwave Background (CMB) \cite{pl}, large-scale structure formation \cite{tegmark}, and gravitational lensing \cite{massey} suggest that dark matter significantly outweighs baryonic matter in the cosmos. Given its dominant presence, exploring its impact on dense astrophysical objects is reasonable.
	
	Although the true nature of dark matter remains a mystery, its gravitational effects on astrophysical systems, such as galaxies, galaxy clusters and compact stars provide indirect evidence for its existence and motivate theoretical investigations into its possible interactions with ordinary matter. Presence of dark matter inside compact stars, such as neutron stars \cite{leungn,hc1,hc2,hc3,guha,das,kain,kark,kark2,kark3,kark4,ng,rou,mari,sco,lopes,bhat,cro,miao,ding,ema}, white dwarfs \cite{chan,bell,zink,niu,curtin,acevedo,leung,steigerwald} and some hypothetically theorized stars, namely quark stars \cite{mukho,pan,lopess,marz,yang,yangy,seda}, has been widely studied and has recently emerged as a particularly active and timely area of research in the quest to understand dark matter through astrophysical phenomena. Dark matter particles in the galactic environment may become gravitationally captured by compact stars through scattering interactions with dense baryonic matter inside the stellar interior \cite{goldman}. As compact stars move through regions of ambient dark matter density, such as the galactic halo, the encounter rate with dark matter particles increases, further enhancing the probability of capture \cite{bert}. When a dark matter particle enters a compact star and undergoes a scattering event, it can lose sufficient kinetic energy to fall below the local escape velocity, thus becoming gravitationally bound to the star \cite{kouvaris}. Captured dark matter particles can undergo repeated scatterings within the star until they lose sufficient energy to remain gravitationally bound, progressively accumulating in the stellar interior \cite{kouvaris}. The presence of such a component within compact stars introduces intriguing questions about its impact on the equation of state (EOS), mass-radius relationship and overall structural stability.
	
	While numerous studies have investigated the impact of dark matter on neutron stars, relatively fewer studies have focused on dark matter effects on strange quark stars (SQS). SQSs are hypothetical compact stars composed of de-confined up, down, and strange quarks. These stars, first proposed by Witten \cite{witten}, fundamentally differ from neutron stars as they are bound primarily by the strong interaction rather than neutron degeneracy pressure. The hypothesis that strange quark matter represents the true ground state of hadronic matter \cite{witten,far} has motivated extensive theoretical exploration. Strange stars can exhibit higher compactness and smaller radii than typical neutron stars. They may explain some observed compact objects whose properties cannot be reconciled with purely hadronic equations of state. Recent astrophysical observations have provided strong motivation for studying strange stars. Compact objects such as SAX J1808.4-3658 \cite{Li}, RX J1856.5-3754 (later declared as neutron stars) \cite{hend}, HER-X1 \cite{Li2}, HESS J1731-347 \cite{hovarth} displayed mass-radius relations that were difficult to explain within conventional neutron star models. The strange star candidates exhibit minimal radii or high compactness, aligning more naturally with predictions from strange matter EOSs. Such observations underscore the importance of exploring strange quark star configurations in greater detail. 
	
	Dark matter effects in compact stars are studied using two-fluid or single-fluid models. Two-fluid models treat DM and BM separately, interacting only via gravity, allowing DM to form a central core, an extended halo or a coextensive distribution in which both fluids share the same spatial boundary \cite{kark,kark2,kark3,mukho,sandin,ciar,reza,thakur,reza2,gle,sun,x,sm,gold,tolos}. While this treatment is physically interesting and meaningful, the technique introduces additional degrees of freedom such as interface radius, matching pressure at the interface and halo boundary, all of which lack direct observational constraints. Moreover, in such models, the total pressure is not derived from a single thermodynamic potential. Even in regions where both components coexist, the system remains non-barotropic, since the local baryonic and dark matter densities may evolve independently. On the other hand, in a single-fluid scenario, dark matter and ordinary matter are considered as a unified single-fluid system, coupled via a shared EOS that incorporates both gravitational and additional interactions, leading to dark matter being distributed throughout the star and typically softening the EOS \cite{kark3,seda,hc1,guha,pann,rouu,lou,sha1,sha2}.  While such models ensure a single thermodynamic closure, they depend sensitively on unknown coupling constants and assume direct interactions between dark matter and baryons, which are strongly constrained by the lack of any experimental detection of such couplings. It is, therefore, more natural to consider that dark matter interacts only gravitationally with ordinary matter.
	
	In this work, we focus on an intermediate regime in which quark matter and self-interacting bosonic dark matter coexist throughout the stellar interior and extend up to a common radius. The two components remain in hydrostatic equilibrium under the same metric potential $\nu(r)$, and, therefore, satisfy identical specific enthalpy gradients
		\begin{equation}
			\frac{dP_Q}{\rho_Q + P_Q} = \frac{dP_D}{\rho_D + P_D},
		\end{equation}
		expressing mechanical equilibrium within the shared gravitational potential. Although this coextensive configuration has previously been discussed as one of the possible regimes of the general two-fluid model, such treatments continued to evolve the pressures of the two components separately through coupled Tolman-Oppenheimer-Volkoff (TOV) equations. In contrast, the present work reformulates the same physical limit in a thermodynamically closed and barotropic manner, replacing the coupled system with a single effective equilibrium equation derived from a unified potential. Each component is assigned its own EOS. The total pressure is then expressed directly as a function of total energy density by combining the contributions from both the fluids using a fixed local volume fraction $\zeta$, which denotes the frozen fraction of quark matter in the mixture. Allowing $\zeta$ to vary spatially would introduce composition gradients and violate the barotropic relation between total pressure and energy density. Hence, a constant $\zeta$ ensures a well-defined thermodynamic closure. This leads to an effective barotropic equilibrium formulation that allows the system to be treated mathematically through a single set of hydrostatic equations, 
		while still accounting for the distinct physical properties of quark matter and dark matter separately. 
		Unlike previous works, one one solves two coupled TOV equations, 
		here the coextensive limit is reformulated as a thermodynamically closed single equilibrium system derived from a unified energy potential
		\begin{equation}
			\varepsilon_{tot}(\rho;\zeta) = \varepsilon_Q(\rho_Q) + \varepsilon_D\big(\rho_D\big),
		\end{equation}
		which guarantees that the effective equation of state
		\begin{equation}
			P(\rho;\zeta) = P_Q(\rho_Q) + P_D\big(\rho_D\big),
		\end{equation}
	 would ensure thermodynamic consistency. The reason for focusing on the coextensive limit is that the repulsive self-interactions among the bosonic dark matter particles provide additional pressure support, preventing gravitational collapse of dark matter particles into a dense core. Consequently, it is logical to assume dark matter to be distributed throughout the stellar interior with its density profile dynamically determined by the equilibrium conditions rather than forming a compact core or an extended halo. This configuration is further supported by the continuous long-term capture of dark matter particles and the stabilizing influence of self-repulsive interactions.
	 
    Among various phenomenological models proposed to investigate the properties of strange quark stars, the MIT bag model and the Nambu-Jona-Lasinio (NJL) have been the most widely employed methods till date. In the MIT bag model, the EOS for strange quark matter takes a simple linear form expressed as $P_Q=\frac{1}{3} (\rho_Q-4 B)$ \cite{witten}, where $\rho_Q$ is the density, $P_Q$ is the isotropic pressure and $B$ is bag constant. To improve upon this description, Dey \textit{et al.} \cite{dey} proposed a new model in which the quark interactions had been characterized by an inter-quark vector potential arising from gluon exchange and a density-dependent scalar potential responsible for restoring chiral symmetry at high densities. The EOS developed by Dey \textit{et al.} can also be approximated to a linear relation of the form $P_Q=a(\rho_Q-\rho_s)$, where $a$ and $\rho_s$ are two parameters representing the softness of the EOS and the surface density, respectively \cite{gond}. 
    
	In our work, we assume the dark matter to be self-repulsive bosonic matter with EOS \cite{kark,kark2,kark3,colpi} $$ P_D=\frac{m_\chi^4}{9 \lambda}(\sqrt{1+\frac{3\lambda \rho_{D}}{m_\chi^4}}-1)^2,$$ where $ m_\chi$ is the dark matter particle mass and $\lambda$ is the coupling constant of self-interaction. Admixing self-interacting bosonic dark matter into strange stars can significantly alter the effective equation of state, leading to distinct compactness or mass-radius relationships. Investigating dark matter admixed strange stars provides a more comprehensive and realistic framework, connecting compact star astrophysics with the properties of dark matter itself. Since both the EOS and the M-R relationship are sensitive to the underlying particle content and interaction dynamics, even a small admixture of dark matter could manifest as measurable shifts in these quantities. This opens up a compelling possibility as the presence of dark matter alters the EOS and it is expected that such modifications might induce distinctive features in the estimation of mass, radius and tidal deformability of a compact star. Precise astrophysical observations, such as NICER data or gravitational wave detectors like LIGO/Virgo can then be utilized to constrain the dark matter content at the interior of such an object. Motivated by this prospect, in this work, we investigate whether observational deviations in the EOS and $M-R$ profiles can serve as indirect signatures of dark matter inside strange stars, thereby providing an astrophysically grounded approach to exploring dark matter properties.
	
	The novelty of the present submission lies in identifying the coextensive ($R_Q = R_D$) configuration as a distinct, physically meaningful limit of the general two-fluid formalism, and in demonstrating that such a configuration admits a thermodynamically closed barotropic equation of state derived from a unified potential. Unlike earlier studies, where one solves the coupled TOV equations for quark and dark components, the present model reformulates the same physical system through a single equilibrium condition based on mechanical coupling and a fixed local composition parameter $\zeta$. Subsequently, the approach provides three distinct gains:
		\begin{enumerate}
			\item it eliminates the need for uncertain interface and boundary conditions inherent to core-halo models, 
			\item It ensures thermodynamic consistency, linking both sectors through an effective chemical potential $\mu_{eff}$, and
			\item it establishes a transparent connection between dark-matter microphysics ($m_{\chi},\lambda,\zeta$) and macroscopic observables ($M,R,z$).
		\end{enumerate}
		Thus, the framework serves as a physically coherent, computationally efficient and conceptually unified limit of two-fluid models, 
		offering an analytic baseline for comparison with full numerical simulations.
		
	The paper is organized as follows: In section \ref{sec2}, we consider a two-fluid system with quark matter following a linear equation of state (EOS) and dark matter described by a self-repulsive bosonic EOS. The system is reduced to a barotropic equilibrium model by defining an effective EOS $P(\rho)$ derived from a unified thermodynamic potential, thereby ensuring thermodynamic consistency. We solve the TOV equation with the effective EOS. In section \ref{sec3}, we obtain the $M-R$ relationship, energy density, pressure and EOS profiles for different quark matter parameters, dark matter fractions, mass of the dark matter particle and the coupling constant of self-interaction for bosonic dark matter. We discuss our results and its observational consequences. In section \ref{sec4}, we conclude by summarizing of our results and also outlining the possible avenues for future exploration.

	\section{Theoretical framework}\label{sec2}
	\subsection{Two-fluid system in a common gravitational field}\label{sec21}
	We consider a static, spherically symmetric spacetime
	\begin{equation}
		ds^{2} = -e^{\nu(r)} dt^{2} + e^{\lambda(r)} dr^{2} + r^{2} d\Omega^{2},
	\end{equation}
	containing two non-interacting perfect-fluid components: strange quark matter (Q) and self-interacting bosonic dark matter (D). 
	The energy-momentum tensor for each component
	\begin{equation}
		T^{(i)}_{\mu\nu} = \left( \rho_{i} + P_{i} \right) u_{\mu} u_{\nu} + P_{i} g_{\mu\nu}, \quad i = Q, D,
	\end{equation}
	which separately satisfy
	\begin{equation}
		\nabla_{\mu} T^{(i)\mu}_{\ \ \ \nu} = 0.
	\end{equation}
	For a static configuration, with $u^{\mu} = e^{-\nu/2}\delta^{\mu}_{0}$, this yields
	\begin{equation}
		\frac{dP_{i}}{dr} = -(\rho_{i} + P_{i})\,\frac{d\nu}{dr}\,\frac{1}{2}, \quad i = Q, D.
		\label{2.1}
	\end{equation}
	
 The total stress-energy tensor takes the form
	\begin{equation}
		T_{\mu\nu} = T^{(Q)}_{\mu\nu} + T^{(D)}_{\mu\nu}.
	\end{equation}
	Dividing Eq.~(\ref{2.1}) by $(\rho_{i} + P_{i})$ gives
	\begin{equation}
		\frac{dP_{i}}{\rho_{i} + P_{i}} = -\frac{1}{2}\, d\nu,
		\label{2.3}
	\end{equation}
 and hence
	\begin{equation}
		\frac{dP_{Q}}{\rho_{Q} + P_{Q}} 
		= \frac{dP_{D}}{\rho_{D} + P_{D}} 
		= -\frac{1}{2}\, d\nu.
		\label{2.4}
	\end{equation}
Note that Eq.~(\ref{2.4}) expresses the equality of specific enthalpy gradients and not the equality of the slopes $dP_{i}/dr$). 
	With the standard definition
	\begin{equation}
		h_{i}(P_{i}) = \int_{0}^{P_{i}} 
		\frac{dP'}{\rho_{i}(P') + P'},
		\label{2.5}
	\end{equation}
	Eq.~(\ref{2.4}) implies 
	\begin{equation}
		dh_{i} = -\frac{1}{2}\, d\nu,
	\end{equation}
meaning each component's enthalpy tracks the same gravitational potentials.

	In a general two-fluid star, $P_{Q}(r)$ and $P_{D}(r)$ may still vanish at different radii (core--halo configuration). 
	To obtain a single common surface, we now impose a composition closure.
	\\
	\subsection{Coextensive limit and fixed local volume fraction}\label{sec22}
	We assume the coextensive limit where $\zeta$ fraction of the total density comes from the quark matter and the rest is the dark matter  contribution
	\begin{equation}
		\rho_{Q} = \zeta\,\rho, 
		\qquad 
		\rho_{D} = (1 - \zeta)\,\rho,
		\qquad 
		0 \leq \zeta \leq 1,
		\label{2.6}
	\end{equation}
	with $\rho = \rho_{Q} + \rho_{D}$.\\
	Physically, this represents a frozen mixture and mathematically, the constancy of the local volume fraction $\zeta$ serves as the closure condition that preserves barotropy of the system. 
	If $\zeta$ is allowed to vary with $r$, an additional term 
	$\rho^{2}(\frac{d\varepsilon_{Q}}{d\rho_Q} - \frac{d\varepsilon_{D}}{d\rho_D})\, (d\zeta/d\rho)$ appears in the pressure term, making $P$ to depend on both $\rho (r)$ and $\zeta(r)$, and thereby, breaks the barotropic relation. Hence, in our analysis, we consider  $\zeta$ to be independent of $r$. However, both $P_{Q}(r)$ and $P_{D}(r)$ are monotonic functions of $r$ and the surface boundary condition on the total pressure for a static star
	\begin{equation}
		P(R) = 0,
		\label{2.8}
	\end{equation}
in our case implies
		\begin{equation}
		P_{Q}(R) = P_{D}(R) = 0, 
		\qquad 
		R_{Q} = R_{D} = R.
		\label{2.9}
	\end{equation}
	Thus, the constant local volume fraction enforces a single stellar boundary in the coextensive limit, although the spatial profiles of strange and dark matter may differ.
	
	\subsection{Unified thermodynamic potential and thermodynamic consistency}\label{sec23}
	Each component satisfies the zero temperature identity  
	\begin{equation}
		P_i = \rho_i \frac{d\varepsilon_i}{d\rho_i} - \varepsilon_i.
		\label{eq:zerotemp}
	\end{equation}
	
	We define the unified potential function as  
	\begin{eqnarray}
		\varepsilon_{tot}(\rho; \zeta)
		&=& \varepsilon_Q(\rho_Q)
		+ \varepsilon_D(\rho_D) \nonumber\\
		&=& \varepsilon_Q(\zeta\rho)
		+ \varepsilon_D\big((1-\zeta)\rho\big),
		\label{eq:etot}
	\end{eqnarray}
	and obtain the total pressure 
	\begin{equation}
		P = \rho\frac{d\varepsilon_{tot}}{d\rho}- \varepsilon_{tot},
		\label{eq:peff1}
	\end{equation}
	yielding the additive form  
	\begin{eqnarray}
		P(\rho)
		= P_Q(\rho_Q)
		+ P_D\big(\rho_D\big)
		=P_Q(\zeta\rho)
		+ P_D\big((1-\zeta)\rho\big).
		\label{eq:peff2}
	\end{eqnarray}
	The above condition guarantees that the effective equation of state is barotropic and that both the pressure and energy density are derived from the same unified potential $\varepsilon_{tot}(\rho;\zeta)$.
	
	\subsection{Equation of state (EOS)}
	
	\subsubsection{Quark matter EOS:}
	We choose a linear quark matter EOS \cite{gond} given by
	\begin{equation}
		P_Q=a  (\rho_Q-\rho_0).  \label{eosf}
	\end{equation}
	For a pure SQS, $\rho_0$ is the surface density and $a=c_s^2$ is the speed of the sound squared in the medium. For physical viability, $0<a<1$. Note that this EOS was earlier used by Sharma {\em et al.} \cite{Sharma} to model relativistic ultra-compact stars.
	Different theoretical models point to the possibility that quark matter inside compact stars may undergo a transition into color-superconducting states, including the  2-flavour superconducting (2SC) and color-flavour locked (CFL) phases \cite{alford,shov}. When the parameter $a$ is varied between 0.2 and 0.8, it is found that the 2SC phase is obtained for $a< 0.33$, whereas the CFL phase corresponds to $a> 0.35$, depending on the NJL parametrization \cite{zdunik}. In the case of a pure quark star, focusing on the CFL phase, we consider two sets of intermediate parameter values: $a=0.4$ , $\rho_0=400 MeV fm^{-3}$ and $a=0.6$ , $\rho_0=250 MeV fm^{-3}$.
	
	\subsubsection{Dark matter EOS:}
	In this study, we chose self-repulsive bosonic matter as dark matter to provide additional pressure for the dark matter component to be present throughout the star. We assume that the dark matter component inside a strange star exists at effectively zero temperature. This assumption is physically motivated by several factors. Once dark matter particles are gravitationally captured by the dense strange quark matter, they rapidly lose their kinetic energy through repeated scatterings and thermalize to the ambient temperature of the star as discussed before. The temperature of strange stars is negligible ($\sim 10^8 K$ \cite{and}) compared to the rest mass energy of typical dark matter particles, whose mass is expected to be around a few hundred MeV to GeV range. The ratio $\frac{k_BT}{m_{\chi} c^2}$, even at the stellar surface is extremely small, ensuring that dark matter behaves as a cold, non-relativistic fluid throughout the star. This consideration justifies treating the dark matter as fully condensed and neglecting any thermal fluctuations. This is essential for modelling the dark matter as a coherent scalar field forming a Bose-Einstein condensate (BEC) at absolute zero temperature \cite{kark,kark2,kark3}, which can be consistently described within the mean-field approximation. The resulting EOS for bosonic matter with repulsive self-interaction was calculated from a quartic potential $V=\frac{\lambda|{\phi|^4}}{4}$ by Colpi \textit{et al.} (1986) and later utilized by Karkevandi \textit{et al.} (2022) in the context of dark matter admixed neutron stars. In the strong coupling regime, the system can be approximated as a perfect fluid, and the anisotropy of pressure will be ignored, so one can reach the EOS of the self-repulsive bosonic DM as \cite{colpi,kark}:\\
	\begin{equation}
		P_D=\frac{m_\chi^4}{9 \lambda}(\sqrt{1+\frac{3\lambda \rho_{D}}{m_\chi^4}}-1)^2, \label{eosb}
	\end{equation}
	where, where $m_\chi$ is the mass of the bosonic particle and $\lambda$ is the dimensionless coupling constant of self-interaction. The maximum mass of such a pure bosonic star is found to be \cite{ph,colpi,mase},\\
	$M_{max}^{BS}=0.06 \lambda^{1/2} M_{Ch}$=$10 M_\odot\lambda^{1/2}(\frac{100~ MeV}{m_\chi})^2$,
	where $M_{Ch}$ is the Chandrasekhar mass.
	
	\subsubsection{Effective EOS of hybrid quark matter-dark matter system:}
	With these EOSs for the two components, using Eqs.~(\ref{eosf}) and (\ref{eosb}), we write the effective EOS of the hybrid quark matter-dark matter system as
	\begin{equation}
		P= a\zeta (\rho-\rho_0^t)+\frac{m_\chi^4}{9 \lambda}(\sqrt{1+\frac{3\lambda (1-\zeta)\rho}{m_\chi^4}}-1)^2, \label{EOSt}
	\end{equation}
	where $\rho_0^t=\rho_0/\zeta$ is a constant. Obviously, $\zeta=0$ implies pure bosonic dark matter star and $\zeta=1$ constitutes a pure SQS.
	
	It is to be noted that, in the quark EOS, $P_Q = a\,(\rho_Q - \rho_0)$, the surface is defined by $P_Q(R) = 0$. This implies, $\rho_Q(R) = \rho_0$ is surface density for a pure quark star ($\zeta = 1$). In the mixed configuration, the boundary condition on the effective pressure $P(R) = 0$ leads to  $\rho(R) \ge \rho_0$, with equality holding when $\zeta = 1$. Therefore, $\rho_0$ is an intrinsic EOS parameter of the quark sector, rather than the physical surface density of the hybrid star.  
	
	\subsection{TOV equation}\label{sec25}
	For the effective pressure $P(\rho)$ 
	from Eq.~(\ref{EOSt}), the stellar structure follows the usual TOV system for the effective variables:
	\begin{equation}
		\frac{dP}{dr}
		= -\,\frac{(\rho + P)\,\left[M + 4\pi r^3 P\right]}
		{r\,(r - 2M)}, \label{tov}
	\end{equation}
	\begin{equation}
		\frac{dM}{dr} = 4\pi r^2 \rho,
		\label{m}
	\end{equation}
	with the boundary conditions 
	$M(0) = 0$ and $P(R) = 0$. 
	The coextensive closure (Sec.~\ref{sec22}) guarantees a single stellar surface, i.e. $R_Q = R_D = R.$ We use Eqs.~(\ref{EOSt}), (\ref{tov}) and (\ref{m}) assuming a reasonable range of central density ($\rho (r=0)$) values and fixed values of $\rho_0$ and $a$ for different values of dark matter fraction, particle mass and the coupling constant to obtain the mass-radius relationship of the resultant configuration.
	
	\section{Results and physical analysis}\label{sec3}
\begin{figure}
		\centering
		\begin{minipage}{0.45\textwidth}
			\centering
			\includegraphics[width=1\textwidth]{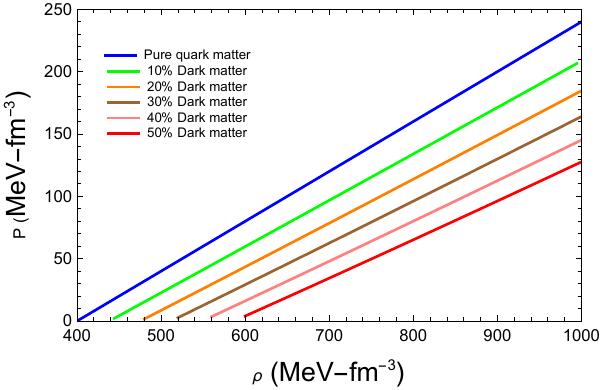}
			\caption{Effective EOS  of dark matter admixed strange stars calculated for various dark matter percentages and fixed values of $\lambda=\pi$ and $m_{\chi}=250~MeV$,$~a=0.4$, $\rho_0=400~ MeV~ fm^{-3}$.}\label{fg1}
		\end{minipage}\hfill
		\begin{minipage}{0.45\textwidth}
			\centering
			\includegraphics[width=1\textwidth]{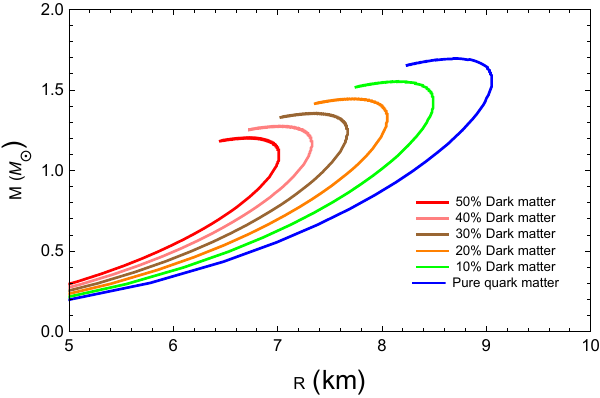}
			\caption{$M-R$ profiles of dark matter admixed strange stars calculated for various dark matter percentages and fixed values of $\lambda=\pi$ and $m_{\chi}=250~MeV$, $~a=0.4$, $\rho_0=400~ MeV~ fm^{-3}$.} \label{fg2}
		\end{minipage}
		\centering
		\begin{minipage}{0.45\textwidth}
			\centering
			\includegraphics[width=1\textwidth]{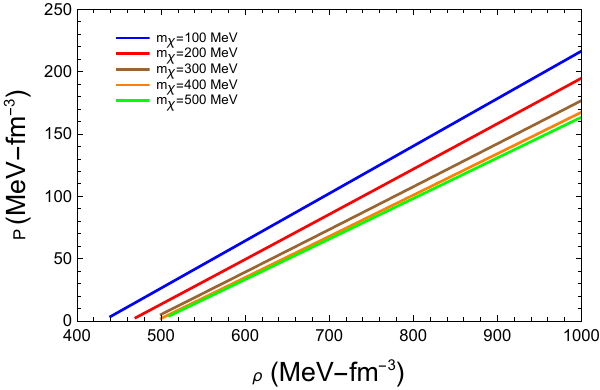}
			\caption{Effective EOS of dark matter admixed strange stars calculated for various dark matter particle masses and fixed values of $\lambda=\pi$ and dark matter percentage $=20\%$,$~a=0.4$, $\rho_0=400~ MeV~ fm^{-3}$.}
			\label{fg3}
		\end{minipage}\hfill
		\begin{minipage}{0.45\textwidth}
			\centering
			\includegraphics[width=1\textwidth]{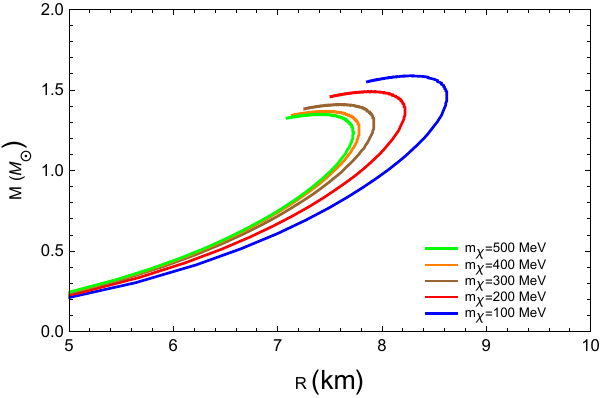}
			\caption{$M-R$ profiles of dark matter admixed strange stars calculated for various dark matter particle masses and fixed values of $\lambda=\pi$ and dark matter percentage $=20\%$, $~a=0.4$, $\rho_0=400~ MeV~ fm^{-3}$.} \label{fg4}
		\end{minipage}
	\end{figure}

	We aim to investigate the impact of dark matter fractions and properties on the pure SQS EOS given by $P=a(\rho-\rho_s)$. For numerical analysis, we assume $a=0.4$ and $\rho_s = 400$~$MeV~fm^{-3}.$ For different volume factions, our results are compiled in Table~\ref{table1}. From Table~\ref{table1} and Fig~\ref{fg1}, it is evident that an increase in dark matter percentage softens the overall EOS. Softening of the EOS means that the pressure increases more slowly with density as dark matter content increases. This results in a less stiff matter distribution, making the star more compressible under gravity, reducing the star's ability to counteract gravitational collapse and leading to smaller maximum mass and radius as shown in Fig~\ref{fg2}. Since the pressure does not support higher masses efficiently, the maximum mass and radius both decrease.
	
	If dark matter behaves as a bosonic condensate, it introduces a self-interacting scalar field that adds an effective pressure component, modifying the EOS. Bosonic dark matter tends to be less stiff than quark matter, which results in lower pressure at a given density. In a SQS, quark matter contributes to degeneracy pressure. Besides this, in advanced models beyond the MIT bag model, e.g., NJL or Dey model \cite{dey}, additional pressure is generated from strong interaction corrections, such as from color superconductivity phases (e.g. 2SC, CFL ). If a fraction of the SQS's mass is replaced by bosonic dark matter, the overall pressure support decreases, softening the EOS.
	
\begin{figure}
		\centering
		\begin{minipage}{0.45\textwidth}
			\centering
			\includegraphics[width=1\textwidth]{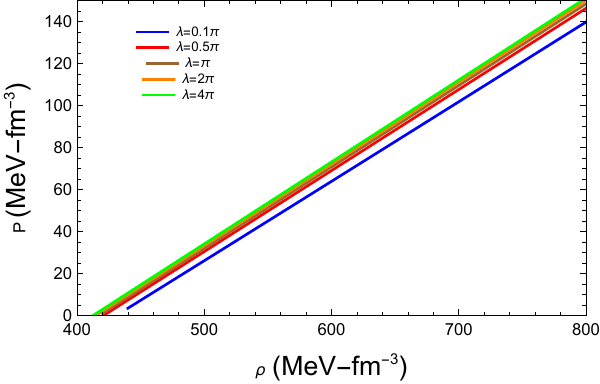}
			\caption{Effective EOS of dark matter admixed strange stars calculated for various values of the self-coupling constant and fixed values of dark matter percentage $=10\%$ and $m_{\chi}=100~MeV$, $~a=0.4$, $\rho_0=400~ MeV~ fm^{-3}$.}\label{fg5}
		\end{minipage}\hfill
		\begin{minipage}{0.45\textwidth}
			\centering
			\includegraphics[width=1\textwidth]{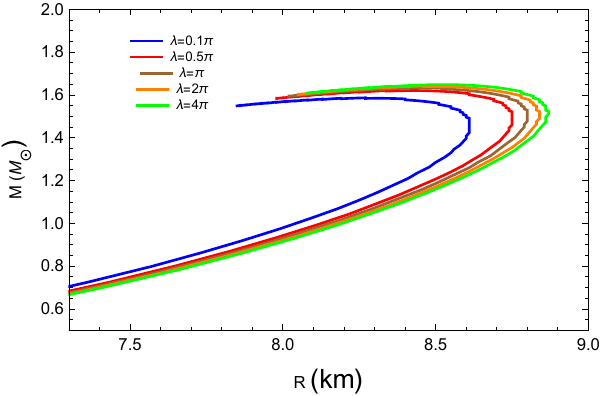}
			\caption{$M-R$ profiles of dark matter admixed strange stars calculated for various values of the self-coupling constant and fixed values of dark matter percentage $=10\%$ and $m_{\chi}=100~MeV$, $~a=0.4$, $\rho_0=400~ MeV~ fm^{-3}$.} \label{fg6}
		\end{minipage}
		\centering
		\begin{minipage}{0.45\textwidth}
			\centering
			\includegraphics[width=1\textwidth]{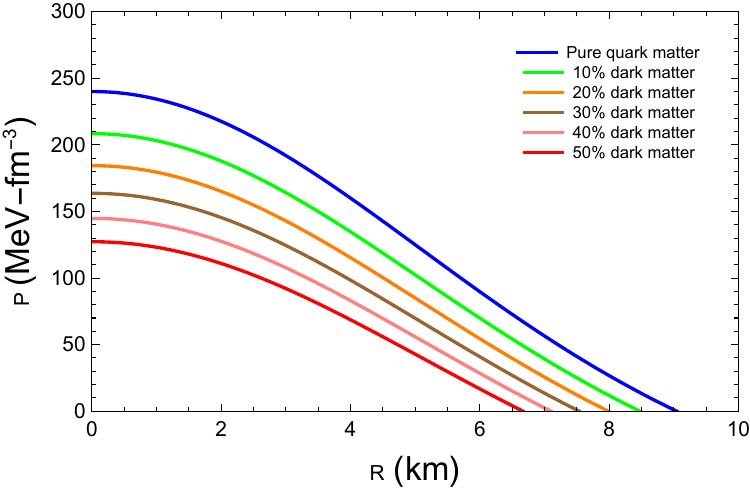}
			\caption{Pressure profiles of dark matter admixed strange stars calculated for various dark matter percentages and fixed values of $\lambda=\pi$ and $m_{\chi}=250~MeV$, $~a=0.4$, $\rho_0=400~ MeV~ fm^{-3}$.} \label{fg7}
		\end{minipage}\hfill
		\begin{minipage}{0.45\textwidth}
			\centering
			\includegraphics[width=1\textwidth]{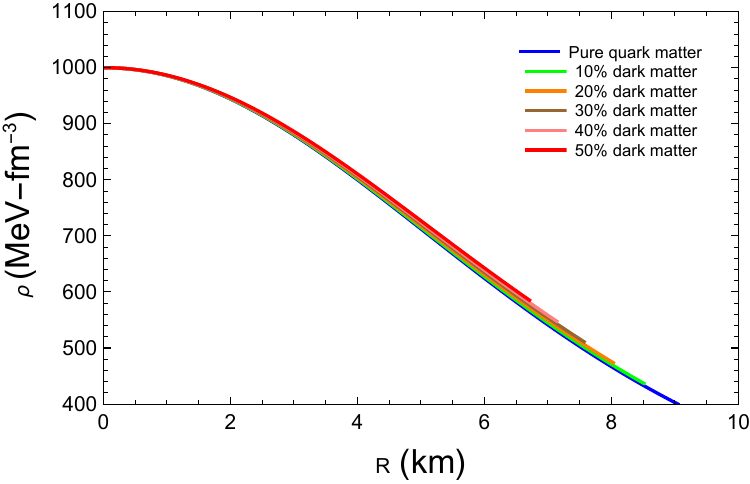}
			\caption{Density profiles of dark matter admixed strange stars calculated for various dark matter percentages and fixed values of $\lambda=\pi$ and $m_{\chi}=250~MeV$, $~a=0.4$, $\rho_0=400~ MeV~ fm^{-3}$.}
			\label{fg8}
		\end{minipage}
	\end{figure}

	It is also to be noted that a reduction in dark matter particle mass and/or an enhancement in the coupling strength contributes to the stiffening of the resultant EOS as shown in Table~\ref{table2} \& Fig~\ref{fg3} and Table~\ref{table3} \& Fig~\ref{fg5}, respectively and hence increasing the maximum mass and radius as shown in Fig~\ref{fg4} and \ref{fg6}. This is because lighter bosons exhibit stronger quantum pressure from the Heisenberg uncertainty principle, which resists gravitational compression more effectively. Simultaneously, stronger self-interactions introduce additional repulsive forces, further increasing pressure at a given energy density. Together, these effects make the bosonic matter more resistant to compression, thereby stiffening the overall EOS. This behavior is consistent with earlier results obtained for a pure bosonic dark star \cite{colpi,mase}.
	
\begin{figure}
		\centering
		\begin{minipage}{0.4\textwidth}
			\centering
			\includegraphics[width=1\textwidth]{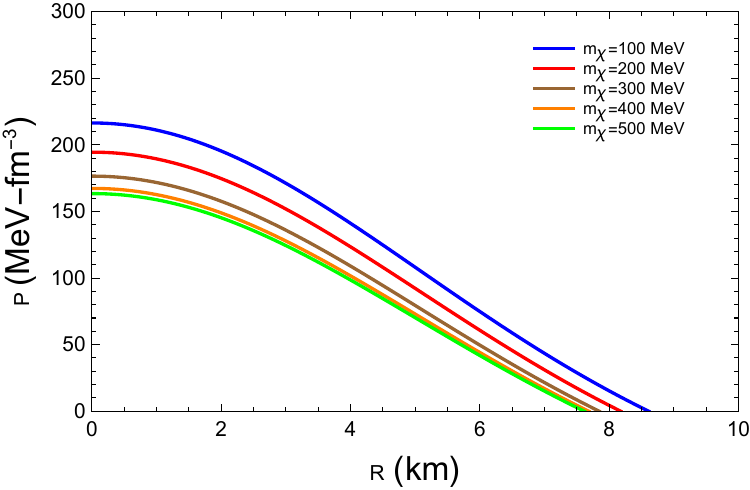}
			\caption{Pressure profiles of dark matter admixed strange stars calculated for various dark matter particle masses and fixed values of $\lambda=\pi$ and dark matter percentage $=20\%$, $~a=0.4$, $\rho_0=400~ MeV~ fm^{-3}$.} \label{fg9}
		\end{minipage}\hfill
		\begin{minipage}{0.4\textwidth}
			\centering
			\includegraphics[width=1\textwidth]{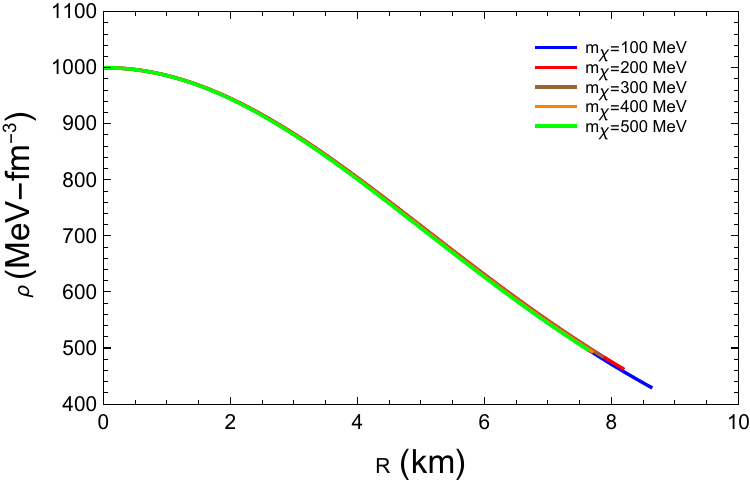}
			\caption{Density profiles of dark matter admixed strange stars calculated for various dark matter particle masses and fixed values of $\lambda=\pi$ and dark matter percentage $=20\%$, $~a=0.4$, $\rho_0=400~ MeV~ fm^{-3}$.}\label{fg10}
		\end{minipage}
		\centering
		\begin{minipage}{0.4\textwidth}
			\centering
			\includegraphics[width=1\textwidth]{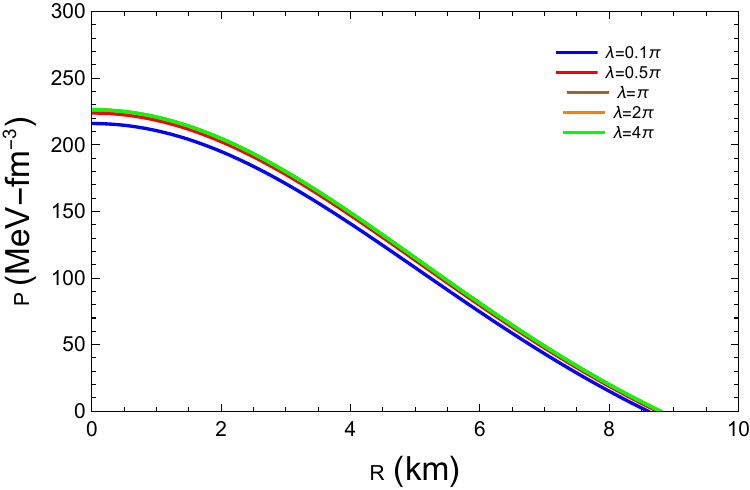}
			\caption{Pressure profiles of dark matter admixed strange stars calculated for various values of the self-coupling constant and fixed values of dark matter percentage $=10\%$ and $m_{\chi}=100~ MeV$, $~a=0.4$, $\rho_0=400~ MeV~ fm^{-3}$.} \label{fg11}
		\end{minipage}\hfill
		\begin{minipage}{0.4\textwidth}
			\centering
			\includegraphics[width=1\textwidth]{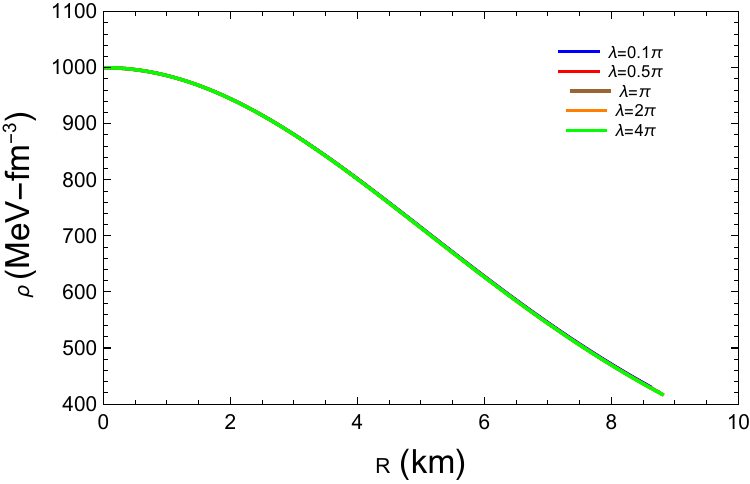}
			\caption{Density profiles of dark matter admixed strange stars calculated for various values of the self-coupling constant and fixed values of dark matter percentage $=10\%$ and $m_{\chi}=100~ MeV$, $~a=0.4$, $\rho_0=400~ MeV~ fm^{-3}$.}
			\label{fg12}
		\end{minipage}
	\end{figure}

	It is noteworthy from Fig.~\ref{fg8}, \ref{fg10} and \ref{fg12}, that the total energy density profiles of the star do not change with any variation, which is expected as the total density always remains the same at any point inside the stellar interior. The only difference is that the surface density is higher for configurations that correspond to softer EOS as the stellar boundary shortens. While some earlier works \cite{kark} indicate a distinctive radial transition between bosonic dark matter and baryonic matter energy density profiles depending on dark matter particle properties, our model does not alter the overall density profile of the star. From Fig~\ref{fg7}, \ref{fg9} and \ref{fg11}, it is evident that for higher dark matter fractions and coupling constants and lower dark matter particle masses, the effective pressure gets reduced.
	\begin{table}
		\centering
		\caption{Variation of the effective EOS with dark matter percentage. 
			Model parameters: $m_\chi = 250~\mathrm{MeV}$, $\lambda = \pi$, $a = 0.4$, $\rho_0 = 400~\mathrm{MeV\,fm^{-3}}$.}
		\label{table1}
		\begin{tabular*}{\columnwidth}{@{\extracolsep{\fill}}ll@{}}
			\hline
			Dark matter percentage & $P~(\mathrm{MeV\,fm^{-3}})$\\
			\hline
			No DM  & $-160.000 + 0.400000\,\rho$\\
			10\%   & $-163.114 + 0.371486\,\rho$\\
			20\%   & $-167.796 + 0.352141\,\rho$\\
			30\%   & $-172.607 + 0.336189\,\rho$\\
			40\%   & $-177.346 + 0.322099\,\rho$\\
			50\%   & $-181.977 + 0.309215\,\rho$\\
			\hline
		\end{tabular*}
	\end{table}
	
	\begin{table}
		\centering
		\caption{Variation of the effective EOS with dark matter particle mass $m_{\chi}$. 
			Model parameters: Dark matter fraction $=0.2$, $\lambda = \pi$, $a = 0.4$, $\rho_0 = 400~\mathrm{MeV\,fm^{-3}}$.}
		\label{table2}
		\begin{tabular*}{\columnwidth}{@{\extracolsep{\fill}}ll@{}}
			\hline
			$m_{\chi}$ (MeV) & $P~(\mathrm{MeV\,fm^{-3}})$\\
			\hline
			100 & $-163.715 + 0.380119\,\rho$\\
			200 & $-167.986 + 0.362471\,\rho$\\
			300 & $-166.592 + 0.343080\,\rho$\\
			400 & $-163.564 + 0.330836\,\rho$\\
			500 & $-161.813 + 0.325193\,\rho$\\
			\hline
		\end{tabular*}
	\end{table}
	
	\begin{table}
		\centering
		\caption{Variation of the effective EOS with coupling constant of dark matter self-interaction $\lambda$. 
			Model parameters: Dark matter fraction $=0.1$, $a = 0.4$, $m_\chi = 100~\mathrm{MeV}$, $\rho_0 = 400~\mathrm{MeV\,fm^{-3}}$.}
		\label{table3}
		\begin{tabular*}{\columnwidth}{@{\extracolsep{\fill}}ll@{}}
			\hline
			$\lambda$ & $P~(\mathrm{MeV\,fm^{-3}})$\\
			\hline
			$0.1\pi$ & $-163.924 + 0.379897\,\rho$\\
			$0.5\pi$ & $-162.964 + 0.386830\,\rho$\\
			$\pi$    & $-162.378 + 0.388689\,\rho$\\
			$2\pi$   & $-161.825 + 0.390019\,\rho$\\
			$4\pi$   & $-161.364 + 0.390974\,\rho$\\
			\hline
		\end{tabular*}
	\end{table}
	
	\begin{table}
		\centering
		\caption{Variation of the maximum mass and corresponding radius with dark matter percentage. 
			Model parameters: $m_\chi = 250~\mathrm{MeV}$, $\lambda = \pi$, $a = 0.4$, $\rho_0 = 400~\mathrm{MeV\,fm^{-3}}$.}
		\label{table4}
		\begin{tabular*}{\columnwidth}{@{\extracolsep{\fill}}lll@{}}
			\hline
			Dark matter percentage & $M_{\mathrm{max}}/M_{\odot}$ & Radius (km)\\
			\hline
			No DM  & 1.6954 & 8.69\\
			10\%   & 1.5528 & 8.15\\
			20\%   & 1.4451 & 7.71\\
			30\%   & 1.3551 & 7.38\\
			40\%   & 1.2752 & 7.04\\
			50\%   & 1.2027 & 6.74\\
			\hline
		\end{tabular*}
	\end{table}
	
	\begin{table}
		\centering
		\caption{Variation of the maximum mass and corresponding radius with dark matter particle mass $m_{\chi}$. 
			Model parameters: Dark matter fraction $=0.2$, $\lambda = \pi$, $a = 0.4$, $\rho_0 = 400~\mathrm{MeV\,fm^{-3}}$.}
		\label{table5}
		\begin{tabular*}{\columnwidth}{@{\extracolsep{\fill}}lll@{}}
			\hline
			$m_{\chi}$ (MeV) & $M_{\mathrm{max}}/M_{\odot}$ & Radius (km)\\
			\hline
			100 & 1.5882 & 8.25\\
			200 & 1.4902 & 7.86\\
			300 & 1.4100 & 7.62\\
			400 & 1.3672 & 7.46\\
			500 & 1.3487 & 7.37\\
			\hline
		\end{tabular*}
	\end{table}
	
	\begin{table}
		\centering
		\caption{Variation of the maximum mass and corresponding radius with the dark matter self-interaction coupling $\lambda$. 
			Model parameters: Dark matter fraction $=0.1$, $a = 0.4$, $m_\chi = 100~\mathrm{MeV}$, $\rho_0 = 400~\mathrm{MeV\,fm^{-3}}$.}
		\label{table6}
		\begin{tabular*}{\columnwidth}{@{\extracolsep{\fill}}lll@{}}
			\hline
			$\lambda$ & $M_{\mathrm{max}}/M_{\odot}$ & Radius (km)\\
			\hline
			$0.1\pi$ & 1.5865 & 8.27\\
			$0.5\pi$ & 1.6203 & 8.42\\
			$\pi$    & 1.6329 & 8.46\\
			$2\pi$   & 1.6414 & 8.46\\
			$4\pi$   & 1.6481 & 8.52\\
			\hline
		\end{tabular*}
	\end{table}

	\subsection{Possible observational consequences:}
	Variation of the maximum mass and corresponding radius with dark matter percentage, dark matter particle mass and coupling constant is shown in Table~\ref{table4}, \ref{table5} and \ref{table6}, respectively. In Table~\ref{table4}, we assume an intermediate mass $m_{\chi}=250~MeV$ for the dark matter particle which is within the range considered earlier in Ref~\cite{lopess} for a bosonic dark matter described by the mean-field Gross-Pitaevskii (GP) EOS \cite{pan,li1,li2,lopesb}. We observe similar results in the maximum mass behavior with increasing dark matter percentage as obtained in Ref~\cite{lopess} for heavier mass dark matter particles. Our results indicate that SQS with higher dark matter content might provide comparatively smaller mass and radius as compared to pure SQS. In other words, a lower mass and radius can be an indication of greater dark matter admixture. 
	
	In the context of stellar objects with comparatively lower radii, it is worth mentioning that Kini \textit{et al.} \cite{kini} recently reported an intriguing mass ($M$) and radius ($R$) measurements for the pulsar XTE~$J1814-338$. Their studies yielded values of the mass of the pulsar XTE $J1814-338$ as $M=1.21_{-0.05}^{+0.05} M_\odot$ and radius $R=7.0_{-0.4}^{+0.4}$ $~km$ at $1\sigma$ confidence level. Such estimates are in agreement with our results for a dark matter admixed star possessing $40\% - 50\%$ dark matter as shown in Fig~\ref{fg2}. For the particular choice of model parameters, we obtain configurations with masses less than $2M_\odot$ and radii less than $9$ $km$. However, higher mass-radius values are also possible by appropriately fixing the values of $a$ and $\rho_0$. 
	
\begin{figure}
		\centering
		\begin{minipage}{0.45\textwidth}
			\centering
			\includegraphics[width=1\textwidth]{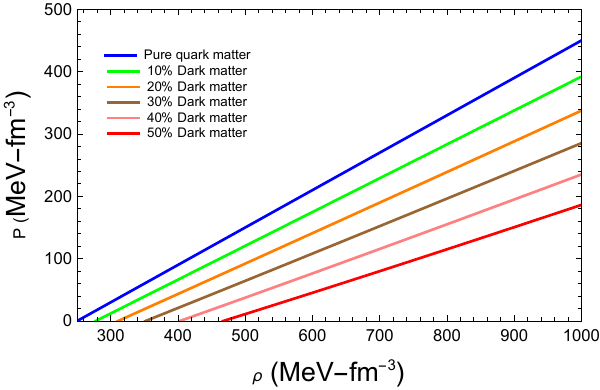}
			\caption{Effective EOS  of dark matter admixed strange stars calculated for various dark matter percentages and fixed values of $\lambda=\pi$ and $m_{\chi}=400~MeV$,$~a=0.6$, $\rho_0=250~ MeV~ fm^{-3}$.}\label{fg13}
		\end{minipage}\hfill
		\begin{minipage}{0.45\textwidth}
			\centering
			\includegraphics[width=1\textwidth]{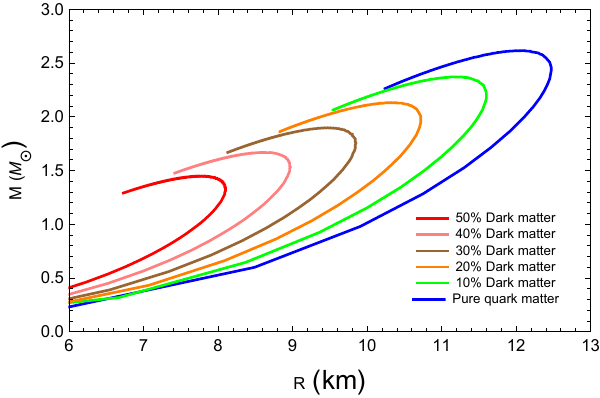}
			\caption{$M-R$ profiles of dark matter admixed strange stars calculated for various dark matter percentages and fixed values of $\lambda=\pi$ and $m_{\chi}=400~MeV$,$~a=0.6$, $\rho_0=250~ MeV~ fm^{-3}$.} \label{fg14}
		\end{minipage}
		\centering
		\begin{minipage}{0.45\textwidth}
			\centering
			\includegraphics[width=1\textwidth]{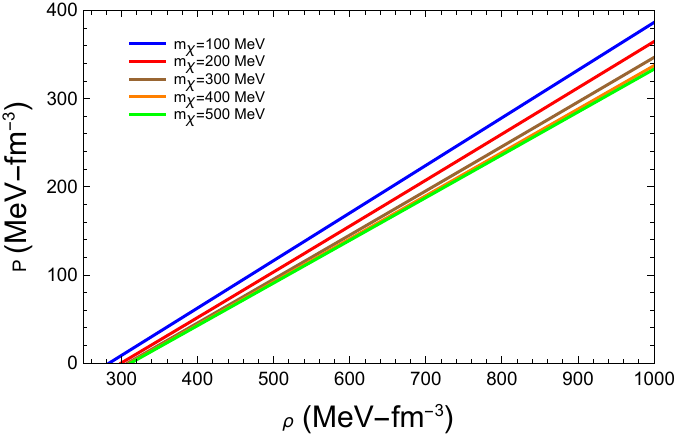}
			\caption{Effective EOS of dark matter admixed strange stars calculated for various dark matter particle masses and fixed values of $\lambda=\pi$ and dark matter percentage $=20\%$, $~a=0.6$, $\rho_0=250~ MeV~ fm^{-3}$.}
			\label{fg15}
		\end{minipage}\hfill
		\begin{minipage}{0.45\textwidth}
			\centering
			\includegraphics[width=1\textwidth]{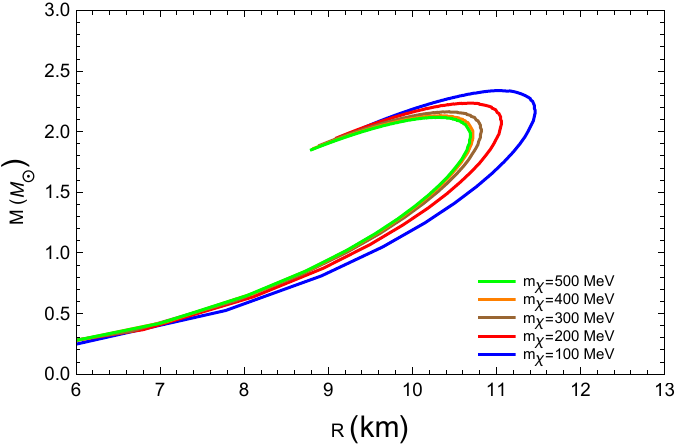}
			\caption{$M-R$ profiles of dark matter admixed strange stars calculated for various dark matter particle masses and fixed values of $\lambda=\pi$ and dark matter percentage $=20\%$,$~a=0.6$, $\rho_0=250~ MeV~ fm^{-3}$.} \label{fg16}
		\end{minipage}
	\end{figure}

\begin{figure}
		\centering
		\begin{minipage}{0.45\textwidth}
			\centering
			\includegraphics[width=1\textwidth]{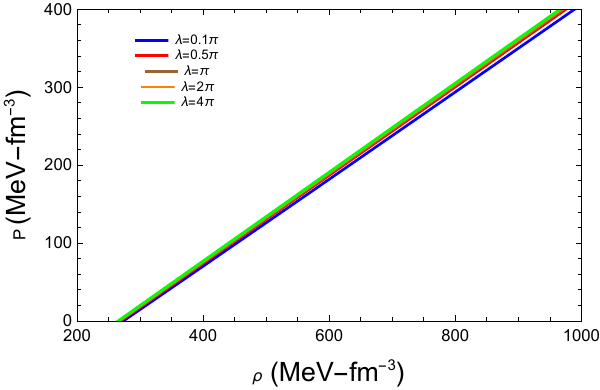}
			\caption{Effective EOS of dark matter admixed strange stars calculated for various values of the self-coupling constant and fixed values of dark matter percentage $=10\%$ and $m_{\chi}=100~MeV$, $~a=0.6$, $\rho_0=250~ MeV~ fm^{-3}$.}\label{fg17}
		\end{minipage}\hfill
		\begin{minipage}{0.45\textwidth}
			\centering
			\includegraphics[width=1\textwidth]{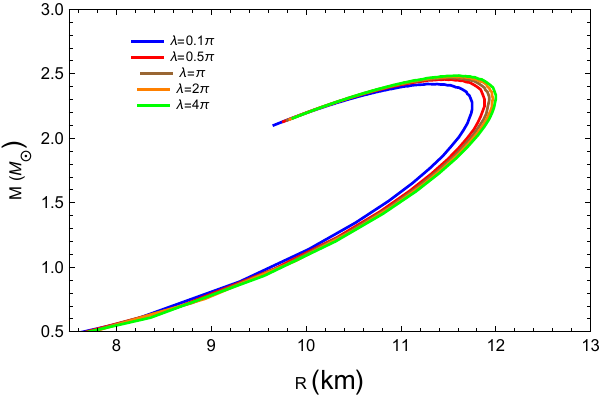}
			\caption{$M-R$ profiles of dark matter admixed strange stars calculated for various values of the self-coupling constant and fixed values of dark matter percentage $=10\%$ and $m_{\chi}=100~MeV$, $~a=0.6$, $\rho_0=250~ MeV~ fm^{-3}$.} \label{fg18}
		\end{minipage}
		\centering
	\end{figure}

	To examine the influence of stiffer quark matter and to account for heavier pulsars with observed masses around or greater than $2\,M_\odot$  such as PSR~$J0348+0432$ ($M = 2.01 \pm 0.04\,M_\odot$~\cite{anto}, $R= 12.246-12.957~\,\mathrm{km}$ (model dependent estimation)~\cite{zhao2016}), PSR~$J0740+6620$ ($M = 2.08 \pm 0.07\,M_\odot$, $R = 13.7^{+2.6}_{-1.5}~\,\mathrm{km}$ (2021)~\cite{miller}, $M = 2.073 \pm 0.069\,M_\odot$, $R = 12.49^{+1.28}_{-0.88}~\,\mathrm{km}$ (2024)~\cite{salmi2024}), PSR~$J0952-0607$ ($M = 2.35 \pm 0.17\,M_\odot$~\cite{romani}, $R = 13.21 \pm 0.96~\,\mathrm{km}$ (model dependent estimation)~\cite{elhanafy2024})), we obtain the $M-R$ relations for $a = 0.6$ and $\rho_0 = 250~\mathrm{MeV\,fm^{-3}}$. This parameter set yields stable configurations with maximum masses exceeding $2\,M_\odot$ as shown in Table~\ref{table10}-\ref{table12}. This demonstrates that the model can accommodate high mass compact stars within a realistic range of quark matter stiffness. This stiffer EOS yields a maximum mass of about maximum mass of about ~$2.6\,M_\odot$ for the pure quark case, placing it in the mass gap region between the heaviest neutron stars and the lightest black holes, observed in gravitational wave event $GW190814$ \cite{abbott}. When dark matter is included, the qualitative behavior of the effective EOS and the $M$–$R$ relation with varying dark-matter fraction, particle mass, and self-interaction strength remains unchanged. That is, increasing dark matter fraction softens the EOS and lowers the mass and radius, whereas lighter particles or stronger self-interaction stiffen it as shown in Table~\ref{table7}-\ref{table9} and Fig~\ref{fg13}-\ref{fg18}. It is to be noted that for our chosen set of parameters, pulsars with higher masses are allowed only within the dark matter fraction range of approximately $0\%-20\%$. When the dark matter contribution increases to about $30\%$, the maximum mass of the configuration drops below $2\,M_{\odot}$.
	
	Variation of the predicted radii with dark matter percentage for various pulsars is summarized in Table \ref{tab:pulsar_dm}. Comparison with observed radii suggests that PSR $J0740+6620$ most likely contains little or no dark matter, as its measured radius aligns closely with the model prediction for the pure quark configuration. In contrast, for XTE~$J1814-338$, a comparatively higher dark matter fraction, around $40-50\%$, is indicated to achieve consistency with the observed radius. However, these results are inherently model-parameter dependent. Varying the stiffness parameter $a$, surface density $\rho_0$ and dark matter parameters changes the predicted radius while preserving the qualitative trends. A general pattern emerges from the analysis. Heavier pulsars with higher observed masses can remain stable only if the dark matter fraction is small. In contrast, ultra compact objects with smaller masses and radii can accommodate a significantly higher dark matter admixture. This division between high mass, low DM and low mass, high DM configurations offers a clear observational diagnostic for future surveys. Overall, the structural behavior reported here remains robust under moderate variations in the underlying quark and dark matter parameters, and in all cases the resulting compactness values remain well within the Buchdahl limit.

	\begin{table}
		\centering
		\caption{Variation of the effective EOS with dark matter percentage. 
			Model parameters: $m_\chi = 400~\mathrm{MeV}$, $\lambda = \pi$, $a = 0.6$, $\rho_0 = 250~\mathrm{MeV\,fm^{-3}}$.}
		\label{table7}
		\begin{tabular*}{\columnwidth}{@{\extracolsep{\fill}}ll@{}}
			\hline
			Dark matter percentage & $P~(\mathrm{MeV\,fm^{-3}})$\\
			\hline
			No DM   & $-150.000 + 0.600000\,\rho$\\
			10\%    & $-150.723 + 0.542656\,\rho$\\
			20\%    & $-152.650 + 0.489699\,\rho$\\
			30\%    & $-155.122 + 0.439614\,\rho$\\
			40\%    & $-159.182 + 0.393304\,\rho$\\
			50\%    & $-163.191 + 0.348357\,\rho$\\
			\hline
		\end{tabular*}
	\end{table}
	
	\begin{table}
		\centering
		\caption{Variation of the effective EOS with dark matter particle mass $m_{\chi}$. 
			Model parameters: Dark matter fraction $=0.2$, $\lambda = \pi$, $a = 0.6$, $\rho_0 = 250~\mathrm{MeV\,fm^{-3}}$.}
		\label{table8}
		\begin{tabular*}{\columnwidth}{@{\extracolsep{\fill}}ll@{}}
			\hline
			$m_{\chi}$ (MeV) & $P~(\mathrm{MeV\,fm^{-3}})$\\
			\hline
			100 & $-153.243 + 0.539515\,\rho$\\
			200 & $-156.496 + 0.520585\,\rho$\\
			300 & $-154.789 + 0.500812\,\rho$\\
			400 & $-152.507 + 0.489511\,\rho$\\
			500 & $-151.230 + 0.484464\,\rho$\\
			\hline
		\end{tabular*}
	\end{table}
	
	\begin{table}
		\centering
		\caption{Variation of the effective EOS with coupling constant of dark matter self-interaction $\lambda$. 
			Model parameters: Dark matter fraction $=0.1$, $a = 0.6$, $m_\chi = 100~\mathrm{MeV}$, $\rho_0 = 250~\mathrm{MeV\,fm^{-3}}$.}
		\label{table9}
		\begin{tabular*}{\columnwidth}{@{\extracolsep{\fill}}ll@{}}
			\hline
			$\lambda$ & $P~(\mathrm{MeV\,fm^{-3}})$\\
			\hline
			$0.1\pi$ & $-152.927 + 0.558606\,\rho$\\
			$0.5\pi$ & $-152.416 + 0.566117\,\rho$\\
			$\pi$    & $-151.975 + 0.568165\,\rho$\\
			$2\pi$   & $-151.545 + 0.569654\,\rho$\\
			$4\pi$   & $-151.172 + 0.570723\,\rho$\\
			\hline
		\end{tabular*}
	\end{table}

	\begin{table}
		\centering
		\caption{Variation of the maximum mass and the corresponding radius with dark matter percentage. 
			Model parameters: $m_\chi = 400~\mathrm{MeV}$, $\lambda = \pi$, $a = 0.6$, $\rho_0 = 250~\mathrm{MeV\,fm^{-3}}$.}
		\label{table10}
		\begin{tabular*}{\columnwidth}{@{\extracolsep{\fill}}lll@{}}
			\hline
			Dark matter percentage & $M_{\mathrm{max}}/M_{\odot}$ & Radius (km)\\
			\hline
			No DM  & 2.6160 & 12.11\\
			10\%   & 2.3716 & 11.25\\
			20\%   & 2.1310 & 10.37\\
			30\%   & 1.8980 & 9.47\\
			40\%   & 1.6680 & 8.70\\
			50\%   & 1.4483 & 7.81\\
			\hline
		\end{tabular*}
	\end{table}
	
	\begin{table}
		\centering
		\caption{Variation of the maximum mass and the corresponding radius with dark matter particle mass $m_{\chi}$. 
			Model parameters: Dark matter fraction $=0.2$, $\lambda = \pi$, $a = 0.6$, $\rho_0 = 250~\mathrm{MeV\,fm^{-3}}$.}
		\label{table11}
		\begin{tabular*}{\columnwidth}{@{\extracolsep{\fill}}lll@{}}
			\hline
			$m_{\chi}$ (MeV) & $M_{\mathrm{max}}/M_{\odot}$ & Radius (km)\\
			\hline
			100 & 2.3387 & 11.03\\
			200 & 2.2353 & 10.69\\
			300 & 2.1650 & 10.44\\
			400 & 2.1329 & 10.34\\
			500 & 2.1189 & 10.26\\
			\hline
		\end{tabular*}
	\end{table}
	
	\begin{table}
		\centering
		\caption{Variation of the maximum mass and the corresponding radius with the dark matter self-interaction coupling $\lambda$. 
			Model parameters: Dark matter fraction $=0.1$, $a = 0.6$, $m_\chi = 100~\mathrm{MeV}$, $\rho_0 = 250~\mathrm{MeV\,fm^{-3}}$.}
		\label{table12}
		\begin{tabular*}{\columnwidth}{@{\extracolsep{\fill}}lll@{}}
			\hline
			$\lambda$ & $M_{\mathrm{max}}/M_{\odot}$ & Radius (km)\\
			\hline
			$0.1\pi$ & 2.4206 & 11.38\\
			$0.5\pi$ & 2.4569 & 11.46\\
			$\pi$    & 2.4690 & 11.52\\
			$2\pi$   & 2.4787 & 11.55\\
			$4\pi$   & 2.4852 & 11.62\\
			\hline
		\end{tabular*}
	\end{table}
	
	\begin{table*}
		\centering
		\caption{Predicted stellar radii for pulsars with varying dark matter fractions.}
		\label{tab:pulsar_dm}
		\setlength{\tabcolsep}{5pt} 
		\renewcommand{\arraystretch}{1.1} 
		\begin{tabular*}{\textwidth}{@{\extracolsep{\fill}}lcccrr@{}}
			\hline
			Star & $M_{\mathrm{obs}}/M_{\odot}$ & $R_{\mathrm{obs}}$ (km) & Model Parameters & DM (\%) & $R_{\mathrm{pred}}$ (km) \\
			\hline
			PSR $J0740+6620$ & $2.073\pm0.069$~\cite{salmi2024} & $12.49^{+1.28}_{-0.88}$~\cite{salmi2024} & $m_\chi = 400~\mathrm{MeV}$, $\lambda = \pi$, $a = 0.6$, $\rho_0 = 250~\mathrm{MeV\,fm^{-3}}$ & 0  & $\approx12.23$ \\
			&                                  &                                         &             & 10 & $\approx11.55$\\
			&                                  &                                         &             & 20 & $\approx10.64$ \\
			\hline
			PSR $J0348+0432$ & $2.01\pm0.04$~\cite{anto} & -- & $m_\chi = 400~\mathrm{MeV}$, $\lambda = \pi$, $a = 0.6$, $\rho_0 = 250~\mathrm{MeV\,fm^{-3}}$ & 0  & $\approx12.10$ \\
			&                          &    &             & 10 & $\approx11.51$ \\
			&                          &    &             & 20 & $\approx10.71$\\
			\hline
			PSR $J0952$-$0607$ & $2.35\pm0.17$~\cite{romani} & -- & $m_\chi = 400~\mathrm{MeV}$, $\lambda = \pi$, $a = 0.6$, $\rho_0 = 250~\mathrm{MeV\,fm^{-3}}$ & 0  & $\approx12.45$ \\
			&                            &    &             & 10 & $\approx11.42$ \\
			\hline
			XTE $J1814-338$ & $1.21\pm0.05$~\cite{kini} & $7.0\pm0.4$~\cite{kini} & $m_\chi = 250~\mathrm{MeV}$, $\lambda = \pi$, $a = 0.4$, $\rho_0 = 400~\mathrm{MeV\,fm^{-3}}$ & 0  & $\approx8.70$ \\
			&                                &                                 &             & 10 & $\approx8.33$ \\
			&                                &                                 &             & 20 & $\approx7.99$ \\
			&                                &                                 &             & 30 & $\approx7.66$ \\
			&                                &                                 &             & 40 & $\approx7.31$ \\
			&                                &                                 &             & 50 & $\approx6.74$ \\
			\hline
		\end{tabular*}
	\end{table*}

	\section{Concluding remarks}\label{sec4}
	In this work, we have investigated the structural and thermodynamic properties of strange quark stars admixed with self-interacting bosonic dark matter under the coextensive equilibrium configuration, where both components coexist throughout the stellar interior and share a common spacetime geometry. The formulation employs a barotropic effective equation of state derived from a unified thermodynamic potential, ensuring that the combined system is thermodynamically consistent. 
		The assumption of a fixed local volume fraction provides a closed equilibrium description while preserving the independent physical identity of the two sectors.
		
	Our analysis shows that the presence of dark matter significantly modifies the global properties of the star. An increase in dark matter fraction systematically softens the effective equation of state, leading to lower maximum masses and smaller radii, whereas lighter dark matter particles or stronger self-interactions produce stiffer configurations. These variations establish a direct connection between dark matter microphysics (through the parameters $(m_\chi,\,\lambda,\,\zeta)$) and observable stellar quantities such as mass, radius and surface redshift. 
		The resulting configurations are consistent with ultra compact X-ray sources like XTE~J1814$-$338, which falls within the predicted range for higher dark-matter fractions. For a stiffer quark matter EOS ($a=0.6$, $\rho_0=250~\mathrm{MeV\,fm^{-3}}$), the model yields maximum masses above $2\,M_\odot$, compatible with massive pulsars such as PSR~J0348+0432, PSR~J0740+6620, and PSR~J0952-0607.
	
	Although the coextensive configuration corresponds to one of the possible limits of the general two-fluid system, the present work reformulates that limit in a conceptually complete and computationally efficient way. In earlier studies, the coextensive case was only treated as a special numerical configuration within coupled TOV systems, where the two pressures evolved independently and the equilibrium was dependent on additional boundary and interface conditions. In our case, by introducing a unified thermodynamic potential and enforcing barotropic closure, we show that this regime can instead be described by a single hydrostatic equilibrium equation derived from a consistent energy functional. This eliminates the need for subsequent matching conditions, provides direct links between microscopic dark-matter parameters and macroscopic observables and allows the configuration to be analyzed in a closed and transparent form. Thus, rather than being a mere simplification, the formulation offers a physically coherent and mathematically tractable description of the mixed-phase configuration.
	
	It is noteworthy that the changes in stiffness and compactness identified in our analysis should be interpreted as macroscopic responses of the composite fluid to variations in dark matter content, rather than microscopic phase transitions within quark matter. The framework, therefore, extends the strange star paradigm into the dark sector in a consistent manner, linking the microphysics of dark matter to measurable stellar observables.
	
	While the present study focuses on equilibrium structure and global stellar properties, the formulation can naturally be extended to the computation of tidal deformability, which depends on the effective equation of state and compactness relations discussed here. Since the tidal deformability is strongly correlated with the stellar radius of a given mass, the softening or stiffening trends noticed in this work can be directly translated into qualitative predictions for the tidal response. A detailed tidal analysis based on the present coextensive model will be carried out in a future work to enable quantitative comparisons with gravitational wave constraints.
	
	It should be stressed that despite many promising prospects in our study, several challenges remain. The precise nature of dark matter interactions with fermionic matter, the stability of dark matter admixed interiors of SQSs, require further probe. Observational data from binary merger experiment should help us in constraining the EOS vis-a-vis matter composition of such class of stars. This may also help us understand the distinguishing features of dark matter admixed neutron/quark stars. In our future works, we aim to refine our dark matter admixed stellar models by incorporating realistic self-interaction terms and exploring its broader astrophysical consequences.
	
	In summary, the present study provides a self-consistent, barotropic and geometrically unified description of dark matter admixed strange stars. It captures the essential physics of quark- dark matter coexistence within a tractable relativistic framework, offering both conceptual clarity and computational efficiency. Our study shows how dark matter parameters such as mass, self-coupling and fractional content can modify the macroscopic behavior of a dark matter admixed strange star. The study also illustrates how different EOS parameters corresponding to the quark matter merely rescale the maximum mass and radius, without altering the fundamental physical behavior of the mixture. Interestingly, a modest dark matter admixture turns out to soften the effective equation of state and shrink the maximum mass limit. In conclusion, our results form a robust theoretical baseline for interpreting dark matter effects on compact stars and opens up the possibility of connecting dark sector microphysics with future multi-messenger astrophysical observations.

	\begin{acknowledgements}
		RS gratefully acknowledges support from the Inter-University Centre for Astronomy and Astrophysics (IUCAA), Pune, India, under its Visiting Research Associateship Programme.
	\end{acknowledgements}
	
		\section*{Declarations}
	
	\begin{itemize}
		\item Funding: Not applicable.
		\item Conflict of interest: The authors declare no conflict of interest.
		\item Ethics approval and consent to participate: All authors have read and agreed to the
		published version of the manuscript.
		\item Data availability: This article’s data is accessible within the public
		domain as specified and duly referenced in the citations.
		\item Materials availability: Not applicable.
		\item Code availability: Not applicable.
		\item Author contribution: All the authors have contributed equally to the manuscript.
	\end{itemize}

\end{document}